\documentclass[12pt,oneside,a4paper]{article}
\usepackage{hyperref}
\hypersetup{colorlinks,bookmarksopen,bookmarksnumbered,citecolor=blue,
linkcolor=black,pdfstartview=FitH,urlcolor=blue}

\usepackage{graphicx}
\usepackage{amssymb}
\usepackage{cite}
\usepackage{bm}
\usepackage{indentfirst}
\usepackage{amsmath}
\usepackage{hhline}
\usepackage{multirow}
\usepackage{color}

\allowdisplaybreaks
\numberwithin{equation}{section}

\renewcommand{\bar}{\overline}

\definecolor{dred}{rgb}{0.7,0.15,0.09}
\definecolor{dblue}{rgb}{0,0.12,0.64}
\definecolor{dgreen}{rgb}{0.2,0.51,0.19}

\begin{document}

\begin{titlepage}

\begin{flushright}
KANAZAWA-21-11
\end{flushright}

\begin{center}

\vspace{1cm}
{\large\textbf{
Distinctive signals of boosted dark matter\\ from its semi-annihilation
}
 }
\vspace{1cm}

\renewcommand{\thefootnote}{\fnsymbol{footnote}}
Takashi Toma$^{1,2}$\footnote[1]{toma@staff.kanazawa-u.ac.jp}
\vspace{5mm}

\textit{
 $^1${\mbox{Institute of Liberal Arts and Science, Kanazawa University, Kanazawa 920-1192, Japan}}\\
 $^2${Institute for Theoretical Physics, Kanazawa University, Kanazawa 920-1192, Japan}
}

\vspace{8mm}

\abstract{
Dark matter can be boosted by various mechanisms, which may produce characteristic signals that are different from those of canonical dark matter. 
We show that the semi-annihilation $\chi\chi\to\bar{\chi}\nu$ produces signals that are distinctive from those of other semi-annihilation and standard dark matter annihilation processes. 
Because the boosted dark matter produced by the semi-annihilation process is regarded as a high-energy neutrino, 
the total flux of the dark matter and the accompanying neutrino produce double peaks in the energy close to the dark matter mass. 
We show that it will be possible to detect both of the particles produced at the Sun using future large volume neutrino detectors such as those of the Deep Underground Neutrino Experiment and Hyper-Kamiokande.
}

\end{center}
\end{titlepage}

\renewcommand{\thefootnote}{\arabic{footnote}}
\newcommand{\bhline}[1]{\noalign{\hrule height #1}}
\newcommand{\bvline}[1]{\vrule width #1}

\setcounter{footnote}{0}

\setcounter{page}{1}
%%%%%%%%%%%%%%%%%%%%%%%%%%%%%%%%%%%%%%

\section{Introduction}
The nature of the dark matter in the universe is still unknown, and its investigation is a primary subject of astro-particle physics. 
Weakly interacting massive particles (WIMPs) are some of the well-motivated dark matter candidates. 
WIMPs were thermally produced in the early universe via sufficient interactions with standard model (SM) particles, and their relic abundance is determined 
by the so-called freeze-out mechanism without dependence on the initial conditions. 
However, no clear evidence for WIMPs has yet been found, and strong constraints are imposed on WIMP models. 
In particular, the current constraints on direct detection experiments are very severe. 
The XENON1T and PandaX-4T collaborations have set upper bounds on the WIMP-nucleon spin-independent (SI) elastic scattering cross section of 
$4.1\times10^{-47}~\mathrm{cm}^2$ and $3.3\times10^{-47}~\mathrm{cm}^2$ for a WIMP mass of $30~\mathrm{GeV}$, respectively~\cite{XENON:2018voc, PandaX-4T:2021bab}. 
One of the ideas for accommodating the strong constraints is to consider velocity-suppressed cross sections 
such as those of pseudo-Nambu-Goldstone boson dark matter~\cite{Gross:2017dan} or fermion dark matter with a pseudo-scalar mediator~\cite{Ipek:2014gua}. 

Another method is to consider semi-annihilating dark matter, which is another kind of WIMP and emerges in models with a non-minimal dark sector. 
In general, semi-annihilations are processes like $\chi\chi\to\bar{\chi}X$ with a (anti-)dark matter particle, $\chi(\bar{\chi})$ 
and SM particle, $X$ in the final state~\cite{Hambye:2008bq, DEramo:2010keq}.\footnote{The corresponding CP conjugate process, $\bar{\chi}\bar{\chi}\to \chi\bar{X}$ also exists.} 
This kind of dark matter is also thermally produced via a freeze-out mechanism just like WIMPs. 
However, it can have relatively weaker couplings with the SM particles compared to WIMPs. 
For example, when an extra $U(1)$ symmetry is spontaneously broken to a $\mathbb{Z}_3$ symmetry, the lightest $\mathbb{Z}_3$ charged particle 
can be a candidate of semi-annihilating dark matter. 
Semi-annihilations possess some interesting features that are different from those of standard annihilations. 
First, the semi-annihilating dark matter cannot be a self-conjugate particle in minimal models because of the charge conservation of the semi-annihilating processes, 
as can be seen from $\chi\chi\to\bar{\chi}X$. 
Second, semi-annihilations produce (semi-)relativistic (anti-)dark matter particles. 
Such boosted dark matter may be detectable using large volume neutrino detectors such as those of Super-Kamiokande (SK)~\cite{Super-Kamiokande:2017dch}, 
IceCube/DeepCore~\cite{IceCube:2015rnn}, and the next generation experiments at Hyper-Kamiokande (HK)~\cite{Hyper-Kamiokande:2018ofw}, PINGU~\cite{IceCube-PINGU:2014okk}, 
the Deep Underground Neutrino Experiment (DUNE)~\cite{DUNE:2020ypp}, and KN3NeT~\cite{KM3Net:2016zxf}. 

\begin{figure}[t]
\begin{center}
\includegraphics[scale=0.65]{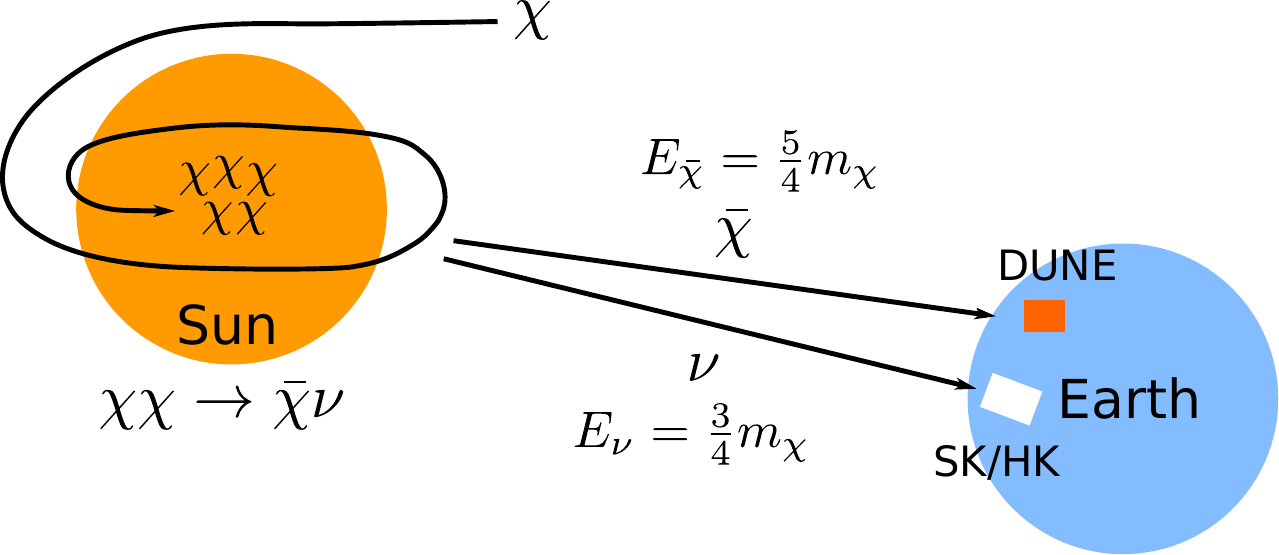}
\caption{Schematic of neutrino and boosted dark matter signals produced by dark matter semi-annihilation at Sun. 
Both particles should be detectable at large volume neutrino detectors such as those at SK/HK and DUNE.}
\label{fig:1}
\end{center}
\end{figure}

In this paper, we consider one of the most native semi-annihilation processes, $\chi\chi\to\bar{\chi}\nu$ where $\nu$ is the light active neutrino in the SM, 
and investigate the simultaneous detection of both particles in the final state at large volume neutrino detectors such as those at SK/HK~\cite{Hyper-Kamiokande:2018ofw} and DUNE~\cite{DUNE:2020ypp}. 
The similar signals produced by the boosted dark matter at the Galactic center and Sun were discussed in refs.~\cite{Agashe:2014yua, Berger:2014sqa, McKeen:2018pbb}. 
Furthermore, the boosted dark matter signals induced by excited dark matter and multi-component dark matter scenarios~\cite{Kong:2014mia, Kopp:2015bfa, Alhazmi:2016qcs, Kim:2016zjx, Aoki:2018gjf, Kim:2019had}, and alternative ideas for boosted dark matter detection~\cite{Kim:2018veo, McKeen:2020vpf} have also been studied. 
On the other hand, this paper focuses on a semi-annihilation process that produces simultaneous signals from the boosted dark matter and high-energy neutrino, 
which are closely correlated with each other. 
This is different from the previous works, and has become a distinctive feature of the semi-annihilation process. 
A schematic of these signals is depicted in Fig.~\ref{fig:1}.

\section{Fluxes from the Sun}
We focus on the neutrino and boosted dark matter fluxes generated by the semi-annihilation process $\chi\chi\to\bar{\chi}\nu$ in the Sun. 
Although it is also possible to consider the fluxes from the galactic center or dwarf spheroidal galaxies, the resultant $\bar{\chi}$ and $\nu$ fluxes are smaller 
than those from the Sun~\cite{Agashe:2014yua, Berger:2014sqa}. 
The energy fluxes of the neutrino and anti-dark matter from the Sun are given by the following:~\cite{Belanger:2013oya, Baratella:2013fya}
\begin{align}
\frac{d\Phi_{\nu/\bar{\chi}}}{dE_{\nu/\bar{\chi}}}=\frac{\Gamma_\mathrm{ann}}{4\pi d_\odot^2}\frac{dN_{\nu/\bar{\chi}}}{dE_{\nu/\bar{\chi}}},
\label{eq:1}
\end{align}
where $d_\odot=1.5\times10^{13}~\mathrm{cm}$ is the distance between the Sun and the Earth, 
$dN_{\nu/\bar{\chi}}/dE_{\nu/\bar{\chi}}$ are the energy spectra of $\nu$ and $\bar{\chi}$, and 
$\Gamma_\mathrm{ann}$ is the annihilation rate, which is proportional to the semi-annihilation cross section times the squared number of accumulated dark matter particles 
in the Sun: $\langle\sigma_{\chi\chi\to\bar{\chi}\nu}{v}\rangle N_\chi^2$. 
The energies of $\nu$ and $\bar{\chi}$  are fixed at $E_{\nu}=3m_\chi/4$ and $E_{\bar{\chi}}=5m_\chi/4$respectively, because the annihilating dark matter particles in the initial state are non-relativistic. 
Thus, the energy spectra $dN_{\nu/\bar{\chi}}/dE_{\nu/\bar{\chi}}$ are simply given by delta functions, and the energy difference between $\nu$ and $\bar{\chi}$ is exactly fixed 
at $E_{\bar{\chi}}-E_\nu = m_\chi/2$. 
The total combined flux would exhibit a double peak structure, which can be a discriminative feature of the semi-annihilation process. 
Note that the CP conjugate process $\bar{\chi}\bar{\chi}\to \chi\bar{\nu}$ should also be taken into account 
if the observed relic abundance is occupied by both $\chi$ and $\bar{\chi}$.

The dark matter particles are captured in the center of the Sun as a result of the energy loss via elastic scattering with nucleons, while 
the number of accumulated anti-dark matter particles decrease via the semi-annihilation. 
Here, we assume that the spin-dependent (SD) elastic scattering cross section $\sigma_\mathrm{SD}$ is dominant over the SI cross section and is velocity-dependent 
($\sigma_\mathrm{SD}\propto v^2$ with dark matter velocity $v$)
because the SI cross section is strongly constrained by dark matter direct detection experiments~\cite{XENON:2018voc, PandaX-4T:2021bab}. 
This velocity dependence leads to significant enhancements of the boosted dark matter signals at the detectors because the typical dark matter local velocity is $v\sim10^{-3}$ 
while it is $v=0.6$ for the boosted dark matter.

The evaporation effect of the accumulated dark matter particles in the Sun could be negligible if the dark matter mass is heavier than $4$ GeV~\cite{Busoni:2013kaa}. 
Then, the capture and semi-annihilation processes reach equilibrium after sufficient time, and the number of dark matter particles in the Sun becomes a constant. 
In this case, the annihilation rate in Eq.~(\ref{eq:1}) can be written in terms of the capture rate, which is proportional to $\sigma_\mathrm{SD}$~\cite{Berger:2014sqa, Garani:2017jcj}. 
We use the analytic fitting function given by
\begin{align}
 \Gamma_\mathrm{ann}\approx&~ 2.6\times10^{21}~[\mathrm{s}^{-1}]\left(\frac{1~\mathrm{TeV}}{m_\chi}\right)^{C_f}\left(\frac{\sigma_\mathrm{SD}}{10^{-40}~\mathrm{cm}^2}\right),
\end{align} 
for the velocity-dependent SD cross section, 
where $C_f=2.15+0.21\log_{10}\left(m_\chi/\mathrm{TeV}\right)$ is the fitting coefficient, and
the above fitting is valid in the mass range $1~\mathrm{GeV}\lesssim m_\chi\lesssim1~\mathrm{TeV}$.
Note that the rate is enhanced by a factor of approximately 20 compared to the constant $\sigma_\mathrm{SD}$~\cite{Berger:2014sqa}. 

The SD elastic scattering cross section is constrained by the direct detection experiments. However it is not as strong as the SI cross section~\cite{XENON:2019rxp, PandaX-II:2018woa}. 
The PICO-60 experiment using 52 kg of C$_3$F$_8$ set the strongest bound for $m_\chi\lesssim100~\mathrm{GeV}$~\cite{PICO:2017tgi}, which was given by 
$3.4\times10^{-41}~\mathrm{cm}^2$ for a $30~\mathrm{GeV}$ WIMP mass. 
For a heavier dark matter mass, the strongest upper bound comes from the neutrino observation at IceCube~\cite{IceCube:2016yoy}. 
The bound depends on the dark matter annihilation channels, and is especially strong for $\tau\bar{\tau}$ and $\nu\bar{\nu}$ channels. 
We adopt the $\nu\bar{\nu}$ bound from ref.~\cite{IceCube:2016yoy}, with a factor of $1/2$ taking into account the fact that only one neutrino is produced at the semi-annihilation. 
This is translated into the upper bound on the annihilation rate, $\Gamma_\mathrm{ann}$. 
Therefore, this implies that the flux of the neutrino and boosted dark matter is also bounded as follows:
\begin{align}
 \Phi_{\nu/\bar{\chi}} \leq 9.2\times10^{-7}~\hspace{-0.05cm}[\mathrm{cm}^{-2}\mathrm{s}^{-1}]\hspace{-0.05cm}
\left(\frac{1~\mathrm{TeV}}{m_\chi}\right)^{C_f}\hspace{-0.15cm}
\left(\hspace{-0.05cm}\frac{\sigma_\mathrm{SD}^\mathrm{exp}}{10^{-40}~\mathrm{cm}^2}\hspace{-0.05cm}\right),
\end{align} 
where $\sigma_\mathrm{SD}^\mathrm{exp}$ is the experimental upper bound of the SD cross section. 

\begin{figure}[t]
\begin{center}
\includegraphics[scale=0.7]{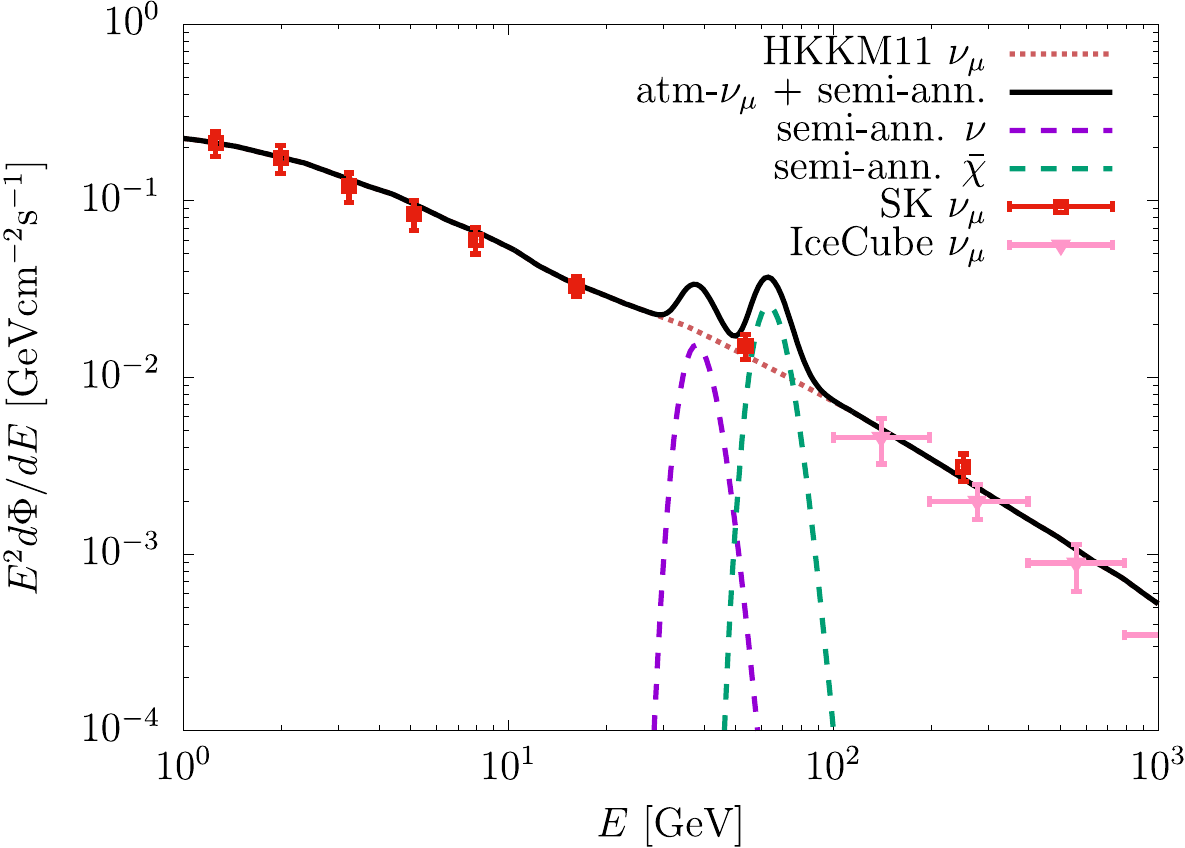}
\caption{Double peak structure of signal fluxes from Sun, where energy resolution of $\Delta E/E=25\%$ is assumed for spectra. 
Values of $m_\chi=50~\mathrm{GeV}$ and $\sigma_\mathrm{SD}=3\times10^{-41}~\mathrm{cm}^2$ were selected for the dark matter mass and SD elastic scattering cross section, respectively, as an example. 
The red and pink points represent the experimental data of SK and IceCube, respectively~\cite{Super-Kamiokande:2015qek}. 
The dark pink dotted line is the fitting with the HKKM11 model~\cite{Honda:2011nf}.}
\label{fig:2}
\end{center}
\end{figure}

An example of the combined flux of $\nu$ and $\bar{\chi}$ is shown in Fig.~\ref{fig:2},
where the delta functions have been filtered with a Gaussian kernel, taking into account the detector energy resolution~\cite{Bertone:2009cb, Ibarra:2012dw}. 
The total flux shows a double peak structure at $E_{\nu}$ and $E_{\bar{\chi}}$. 
Because atmospheric neutrinos can be the primary background for the signals, the total atmospheric muon neutrino fluxes observed at SK and IceCube are also shown in Fig.~\ref{fig:2}. 
Although atmospheric electron neutrinos also exist, the $\nu_e$ flux is much smaller than the muon neutrinos for the energy range $E_{\nu}\gtrsim\mathcal{O}(10)~\mathrm{GeV}$. 

\section{Boosted dark matter signal at DUNE}
DUNE consists of two detectors located at the Fermi National Accelerator Laboratory and Stanford Underground Research Laboratory~\cite{DUNE:2020ypp}. 
The latter uses a 40 kton fiducial mass of liquid argon and can search for the signal from boosted dark matter scattering off a proton 
using a Liquid Argon Time Projection Chamber (LArTPC).\footnote{Although scattering off a neutron can also produce a signal, the event reconstruction is relatively more challenging than that for a proton.}

The expected signal events of the boosted dark matter can be estimated by the following:
\begin{align}
N_\mathrm{sig}=t_\mathrm{exp}\cdot N_p\cdot\Phi_{\bar{\chi}}\cdot\tilde{\sigma}_\mathrm{SD}\cdot\epsilon_\mathrm{eff},
\end{align}
where $t_\mathrm{exp}$ is the exposure time, $N_p$ is the number of target protons at the detector, $\Phi_{\bar{\chi}}$ is the flux from the Sun, 
$\tilde{\sigma}_\mathrm{SD}$ is the scattering cross section of the boosted dark matter with a proton in the argon detector, and $\epsilon_\mathrm{eff}$ is the detector efficiency. 

The elastic scattering cross section for the boosted dark matter ($\tilde{\sigma}_\mathrm{SD}$) is different from 
$\sigma_\mathrm{SD}$ which appears in the above flux calculation. 
Because the SD cross section is assumed to be velocity-suppressed ($\sigma_\mathrm{SD}\propto v^2$) for non-relativistic dark matter, the cross section $\tilde{\sigma}_\mathrm{SD}$ 
is approximately estimated as 
$\tilde{\sigma}_\mathrm{SD}\sim3.6\times10^5\sigma_\mathrm{SD}$ taking into account the velocity difference between the non-relativistic and boosted dark matter.
Note that the effect of the nucleon form factors should also be taken into account for precise calculation, which is expected to induce a $\mathcal{O}(1)$ difference~\cite{Agashe:2014yua, Berger:2014sqa, Berger:2019ttc}. 

From the kinematics of the elastic scattering, the energy of the scattered proton, $E_p$ is in the range of 
\begin{align}
m_p \leq E_p \leq m_p\frac{(E_{\bar{\chi}}+m_p)^2 + E_{\bar{\chi}}^2-m_\chi^2}{(E_{\bar{\chi}}+m_p)^2 - E_{\bar{\chi}}^2+m_\chi^2},
\label{eq:5}
\end{align}
in the rest frame of the proton~\cite{Agashe:2014yua}.
Substituting the energy of the boosted dark matter, $E_{\bar{\chi}}=5m_\chi/4$, the energy of the scattered proton becomes
\begin{align}
m_p\leq E_{p}\lesssim\frac{17}{8}m_{p},
\end{align}
where $m_p\ll m_\chi$ is assumed. 
Thus, the proton energy is less than $2~\mathrm{GeV}$ and almost independent of the dark matter mass. 
In this energy range, resonant scatterings and deep inelastic scatterings can be ignored~\cite{Berger:2019ttc}. 
In addition, this energy is not sufficient to emit Cherenkov light in water and ice at facilities such as SK/HK and IceCube. 
However, it is anticipated to be observed at DUNE because the thresholds of the proton kinetic energy and angular resolution are as low as $50~\mathrm{MeV}$ and $5^\circ$, 
respectively~\cite{DUNE:2015lol}. 
These precise measurements significantly reduce the main background of atmospheric neutrinos~\cite{Agashe:2014yua}. 

The detector efficiency, $\epsilon_\mathrm{eff}$ is estimated to be in the range of $0.80\lesssim \epsilon_\mathrm{eff}\lesssim0.95$ 
for electron neutrinos and muon neutrinos via the charged current interaction when the neutrino energy is in the range of $1~\mathrm{GeV}$ to $2~\mathrm{GeV}$~\cite{DUNE:2015lol}. 
Because there is no $\epsilon_\mathrm{eff}$ estimation for the boosted dark matter, we selected $\epsilon_\mathrm{eff}=0.80$ as a benchmark. 

\begin{figure}[t]
\begin{center}
\includegraphics[scale=0.7]{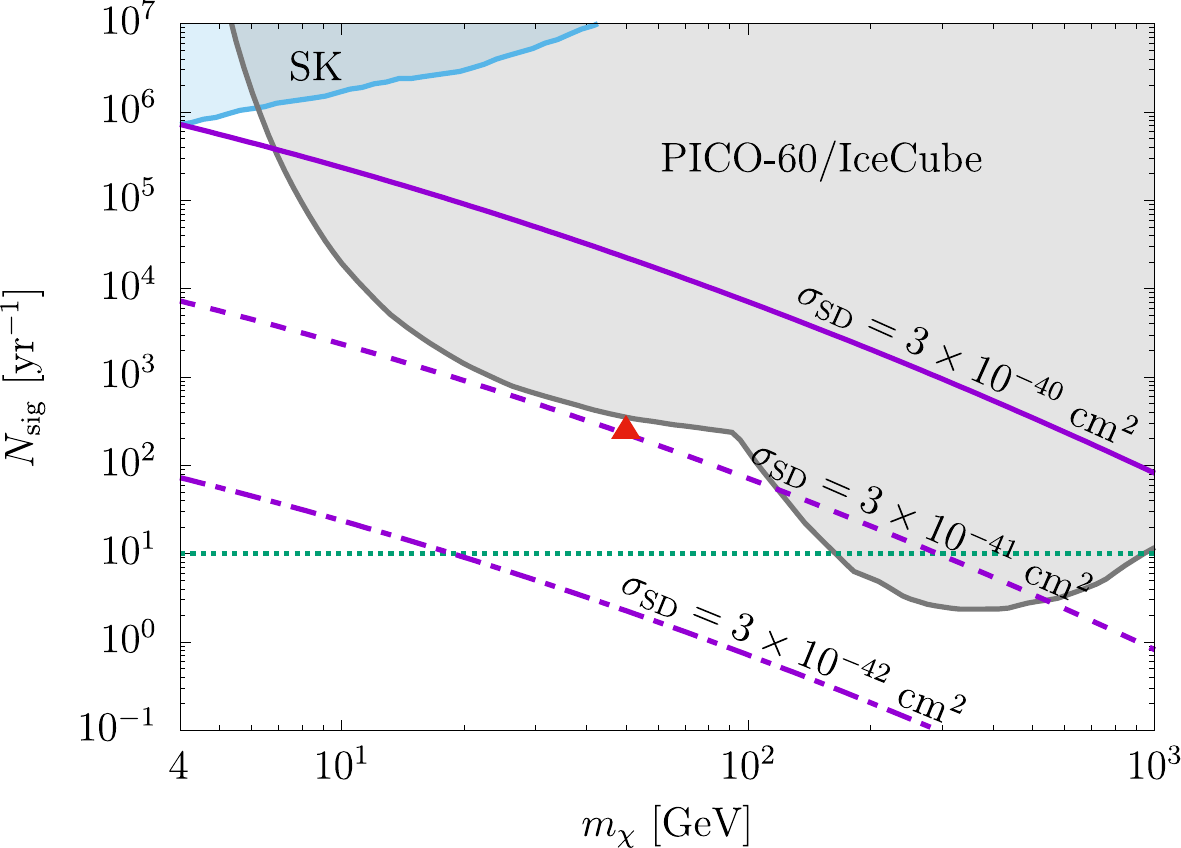}
\caption{Expected number of signal events at DUNE as function of dark matter mass, 
where detector efficiency is taken to be $\epsilon_\mathrm{eff}=0.80$. The red triangle is a sample parameter point used for Fig.~\ref{fig:2}. 
The upper gray region is excluded by PICO-60~\cite{PICO:2017tgi} and IceCube~\cite{IceCube:2016yoy}, 
while the left upper light-blue region is excluded by the observation of atmospheric neutrinos at SK~\cite{Super-Kamiokande:2015qek}.}
\label{fig:3}
\end{center}
\end{figure}

The expected number of signal events per year at DUNE is shown in Fig.~\ref{fig:3}. 
The number of target protons with the 40 kton fiducial mass of liquid argon is $N_p=1.1\times10^{34}$. 
The upper gray region is excluded by the bound of the SD cross section in the non-relativistic limit. 
The left upper light-blue region is also excluded by the observation of the atmospheric neutrinos at SK. 
The SK limit is not as severe because it was evaluated with the atmospheric neutrinos from all directions. 
It is expected that the limit would become an order of magnitude more severe if only the direction of the Sun was taken into account. 
The red triangle is a sample parameter point that was adopted for Fig.~\ref{fig:2}. 
Note that a large number of signals can be expected if the dark matter mass is less than $\mathcal{O}(10)~\mathrm{GeV}$. However, this mass region tends to be excluded by the gamma ray observations via the process $\chi\bar{\chi}\to q\bar{q}$~\cite{MAGIC:2016xys}, which necessarily occurs through the crossing symmetry of the elastic scattering process, $\chi p\to\chi p$. 
The constraint depends on the dark matter models; it can be severe if the annihilation into quarks is $s$-wave, while there is no practical constraint for $p$-wave. 
For the realistic detection of a boosted dark matter signal, a benchmark for the number of events is set to be $10$ events per year, which is shown as a green dotted line in Fig.~\ref{fig:3}. 
As seen in this figure, the dark matter mass should be $m_\chi\lesssim150~\mathrm{GeV}$ so that the required number of 
events are observed.\footnote{Although a dark matter mass that is heavier than $\mathcal{O}(1)~\mathrm{TeV}$ with a large SD cross section, may also produce the detectable number of signals, 
the accompanying neutrino would be outside of the energy threshold of DUNE. 
Even in this case, the neutrino could be detectable at other large volume neutrino detectors such as 
SK/HK~\cite{Super-Kamiokande:2017dch, Hyper-Kamiokande:2018ofw}, IceCube/DeepCore~\cite{IceCube:2015rnn}, PINGU~\cite{IceCube-PINGU:2014okk}, and KM3NeT~\cite{KM3Net:2016zxf}. 
}
The expected number of signal events at the sample parameter point (red triangle) is $318$ per year.
The above calculation is no more than a rough estimation, and a more detailed simulation at DUNE should be performed following refs.~\cite{Berger:2019ttc, DUNE:2020ypp}.
Furthermore, although we have not taken into account the effect of the neutrino oscillations, it should be included for a more sophisticated analysis.

In the case of a constant SD cross section, it is expected that the number of signal events would be smaller than that in the velocity-dependent case. 
Removing the enhancement factors for the $v^2$-dependent SD cross section, the number of signal events becomes 
$0.1$ per year at most when the dark matter mass is $\sim6~\mathrm{GeV}$ as can be seen in Fig.~\ref{fig:3}. 
Therefore, the velocity dependent cross section is essential for the signals of the boosted dark matter. 
Otherwise, a breakthrough in experimental technique is required. 

The accompanying neutrino may also induce another signal, $\nu p\to \nu p$ at a different energy scale. 
Because $E_{\nu}\gg m_p$, deep inelastic scattering dominates in this case. 
The energy range of the scattered proton is given by $m_p\leq E_p\lesssim m_\chi$ from Eq.~(\ref{eq:5}). 
If an experiment at DUNE allows a neutrino signal search in this energy range, searches for the characteristic signals of both the boosted dark matter and neutrinos 
from the semi-annihilation process can be conducted simultaneously, which is the best scenario. 
Even if this is not the case because the neutrino energy is too large (corresponding to a heavy dark matter mass) to be detected, 
a search for the signal from the accompanying neutrinos can be conducted at SK/HK~\cite{Super-Kamiokande:2017dch, Hyper-Kamiokande:2018ofw}. 

\section{Model building}
A velocity-dependent SD cross section was assumed in the above argument. 
Such a cross section can be derived from the anapole moment or scalar-pseudoscalar (SP) interaction given by the following:~\cite{Gelmini:2018ogy}
\begin{align}
\mathcal{L}_\mathrm{ana}&=\frac{1}{\Lambda^2}\bar{\chi}\gamma_{\mu}\gamma_5\partial_{\nu}\chi F^{\mu\nu},\\
\mathcal{L}_\mathrm{SP}&=\frac{1}{\Lambda^2}\left(\bar{\chi}\chi\right)\left(\bar{q}\gamma_5 q\right),
\end{align}
where $\Lambda$ is the theoretical cut-off for the scale, and $F^{\mu\nu}$ is the electromagnetic field strength, which is contracted with the electromagnetic current $eA_{\mu}J^{\mu}$, 
and induces elastic scattering with a proton. 
The resultant SD cross sections from the above interactions are suppressed by the squared velocity ($\sigma_\mathrm{SD}\propto v^2$) for non-relativistic dark matter. 
However, note that for the anapole moment interaction, an SI cross section with the $v^2$ suppression also emerges at the same time, which could be dominant over the SD cross section. 

In addition, the pseudoscalar-pseudoscalar (PP) interaction
\begin{align}
 \mathcal{L}_\mathrm{PP}=\frac{1}{\Lambda^2}\left(\bar{\chi}\gamma_5 \chi\right)\left(\bar{q}\gamma_5 q\right),
\end{align}
induces further suppression of the SD cross section ($\sigma_\mathrm{SD}\propto v^4$). 
In this case, the standard annihilation cross section for $\chi\bar{\chi}\to q\bar{q}$ becomes $s$-wave, and 
the model is strongly constrained by the gamma ray observations of the MAGIC and Fermi-LAT collaborations if the dark matter mass is below $\mathcal{O}(100)~\mathrm{GeV}$~\cite{MAGIC:2016xys}. 

When the velocity-suppression is extremely strong, like the above PP interaction, it is necessary to be careful with loop corrections in the ultra-violet (UV) complete models. 
The loop corrections may induce a new contribution to the cross section that is not velocity-suppressed, and becomes dominant over the velocity-suppressed cross sections, depending on the parameters. 

In the UV complete models with radiative neutrino masses~\cite{Ma:2007gq, Aoki:2014cja} and their further extensions~\cite{Ho:2016aye}, 
the semi-annihilation process $\chi\chi\to \bar{\chi}\nu$ can naturally occur. 
In these models, although neutrino masses at the tree level are forbidden by the imposed $\mathbb{Z}_3$ symmetry, small masses are induced at the two-loop level. 
This symmetry also stabilizes the lightest $\mathbb{Z}_3$ charged fermion and provides a dark matter candidate with the semi-annihilation process explored here.

\section{Summary and discussions}
%summary
We investigated the distinctive boosted dark matter and neutrino signals produced by the semi-annihilation $\chi\chi\to \bar{\chi}\nu$ induced at the Sun. 
These two signals are correlated with each other, and are not produced by the other semi-annihilation and standard dark matter annihilation processes. 
We approximately estimated the boosted dark matter and neutrino fluxes from the Sun, assuming a velocity-suppressed SD cross section. 
Both of the produced boosted dark matter and neutrino signals could be detected at DUNE and SK/HK. 
We also performed a simple estimation of the number of signal events at DUNE, 
and found that the dark matter mass should be $m_\chi\lesssim150~\mathrm{GeV}$ to allow a realistic number of signal events to be observed. 
If these semi-annihilation signals are experimentally confirmed, this will strongly suggest that the dark matter in the universe is a Dirac fermion with spin 1/2. 
This is a unique process that makes it possible to identify the spin of the dark matter particle.

%discussion
Our framework can be applied to other boosted dark matter scenarios such as multi-component dark matter models. 
In this case, the correlation between the two signals would be weaker because the heavier dark matter captured in the Sun and the lighter boosted dark matter 
that could be detectable at DUNE are different particles. 
Furthermore, a similar argument with a velocity-dependent SI cross section could also be studied. 
In this case, the expected $\nu$ and $\bar{\chi}$ fluxes would be smaller than those in the case we have considered because the experimental limit on the SI cross section is much stronger than that of the SD cross section. 
However, at most, a few signal events per year from boosted dark matter can be expected at DUNE if the dark matter mass is less than $10~\mathrm{GeV}$.

\section*{Acknowledgments}
The author would like to thank Mayumi Aoki for carefully reading the manuscript and providing
valuable comments.
This work was supported by a JSPS Grant-in-Aid for Scientific Research (KAKENHI Grant
No. JP20K22349).
The numerical computation in this work was carried out at the
Yukawa Institute Computer Facility.


\begin{thebibliography}{200}
%\cite{XENON:2018voc}
\bibitem{XENON:2018voc}
E.~Aprile \textit{et al.} [XENON],
%``Dark Matter Search Results from a One Ton-Year Exposure of XENON1T,''
Phys. Rev. Lett. \textbf{121}, no.11, 111302 (2018)
%doi:10.1103/PhysRevLett.121.111302
[arXiv:1805.12562 [astro-ph.CO]].
%1274 citations counted in INSPIRE as of 30 Aug 2021

%\cite{PandaX-4T:2021bab}
\bibitem{PandaX-4T:2021bab}
Y.~Meng \textit{et al.} [PandaX-4T
],
%``Dark Matter Search Results from the PandaX-4T Commissioning Run,''
[arXiv:2107.13438 [hep-ex]].
%5 citations counted in INSPIRE as of 30 Aug 2021


%\cite{Gross:2017dan}
\bibitem{Gross:2017dan}
C.~Gross, O.~Lebedev and T.~Toma,
%``Cancellation Mechanism for Dark-Matter\textendash{}Nucleon Interaction,''
Phys. Rev. Lett. \textbf{119}, no.19, 191801 (2017)
%doi:10.1103/PhysRevLett.119.191801
[arXiv:1708.02253 [hep-ph]].
%65 citations counted in INSPIRE as of 02 Sep 2021

%\cite{Ipek:2014gua}
\bibitem{Ipek:2014gua}
S.~Ipek, D.~McKeen and A.~E.~Nelson,
%``A Renormalizable Model for the Galactic Center Gamma Ray Excess from Dark Matter Annihilation,''
Phys. Rev. D \textbf{90}, no.5, 055021 (2014)
%doi:10.1103/PhysRevD.90.055021
[arXiv:1404.3716 [hep-ph]].
%198 citations counted in INSPIRE as of 09 Sep 2021



%\cite{Hambye:2008bq}
\bibitem{Hambye:2008bq}
T.~Hambye,
%``Hidden vector dark matter,''
JHEP \textbf{01}, 028 (2009)
%doi:10.1088/1126-6708/2009/01/028
[arXiv:0811.0172 [hep-ph]].
%247 citations counted in INSPIRE as of 02 Jan 2022

%\cite{DEramo:2010keq}
\bibitem{DEramo:2010keq}
F.~D'Eramo and J.~Thaler,
%``Semi-annihilation of Dark Matter,''
JHEP \textbf{06}, 109 (2010)
%doi:10.1007/JHEP06(2010)109
[arXiv:1003.5912 [hep-ph]].
%154 citations counted in INSPIRE as of 02 Jan 2022


%\cite{Super-Kamiokande:2017dch}
\bibitem{Super-Kamiokande:2017dch}
C.~Kachulis \textit{et al.} [Super-Kamiokande],
%``Search for Boosted Dark Matter Interacting With Electrons in Super-Kamiokande,''
Phys. Rev. Lett. \textbf{120}, no.22, 221301 (2018)
%doi:10.1103/PhysRevLett.120.221301
[arXiv:1711.05278 [hep-ex]].
%33 citations counted in INSPIRE as of 11 Sep 2021

%\cite{IceCube:2015rnn}
\bibitem{IceCube:2015rnn}
M.~G.~Aartsen \textit{et al.} [IceCube],
%``Search for Dark Matter Annihilation in the Galactic Center with IceCube-79,''
Eur. Phys. J. C \textbf{75}, no.10, 492 (2015)
%doi:10.1140/epjc/s10052-015-3713-1
[arXiv:1505.07259 [astro-ph.HE]].
%87 citations counted in INSPIRE as of 05 Jan 2022

%\cite{Hyper-Kamiokande:2018ofw}
\bibitem{Hyper-Kamiokande:2018ofw}
K.~Abe \textit{et al.} [Hyper-Kamiokande],
%``Hyper-Kamiokande Design Report,''
[arXiv:1805.04163 [physics.ins-det]].
%359 citations counted in INSPIRE as of 11 Sep 2021

%\cite{IceCube-PINGU:2014okk}
\bibitem{IceCube-PINGU:2014okk}
M.~G.~Aartsen \textit{et al.} [IceCube-PINGU],
%``Letter of Intent: The Precision IceCube Next Generation Upgrade (PINGU),''
[arXiv:1401.2046 [physics.ins-det]].
%306 citations counted in INSPIRE as of 11 Sep 2021

%\cite{DUNE:2020ypp}
\bibitem{DUNE:2020ypp}
B.~Abi \textit{et al.} [DUNE],
%``Deep Underground Neutrino Experiment (DUNE), Far Detector Technical Design Report, Volume II: DUNE Physics,''
[arXiv:2002.03005 [hep-ex]].
%148 citations counted in INSPIRE as of 06 Sep 2021

%\cite{KM3Net:2016zxf}
\bibitem{KM3Net:2016zxf}
S.~Adrian-Martinez \textit{et al.} [KM3Net],
%``Letter of intent for KM3NeT 2.0,''
J. Phys. G \textbf{43}, no.8, 084001 (2016)
%doi:10.1088/0954-3899/43/8/084001
[arXiv:1601.07459 [astro-ph.IM]].
%544 citations counted in INSPIRE as of 30 Aug 2021


%\cite{Agashe:2014yua}
\bibitem{Agashe:2014yua}
K.~Agashe, Y.~Cui, L.~Necib and J.~Thaler,
%``(In)direct Detection of Boosted Dark Matter,''
JCAP \textbf{10}, 062 (2014)
%doi:10.1088/1475-7516/2014/10/062
[arXiv:1405.7370 [hep-ph]].
%120 citations counted in INSPIRE as of 30 Aug 2021



%\cite{Berger:2014sqa}
\bibitem{Berger:2014sqa}
J.~Berger, Y.~Cui and Y.~Zhao,
%``Detecting Boosted Dark Matter from the Sun with Large Volume Neutrino Detectors,''
JCAP \textbf{02}, 005 (2015)
%doi:10.1088/1475-7516/2015/02/005
[arXiv:1410.2246 [hep-ph]].
%66 citations counted in INSPIRE as of 30 Aug 2021

%\cite{McKeen:2018pbb}
\bibitem{McKeen:2018pbb}
D.~McKeen and N.~Raj,
%``Monochromatic dark neutrinos and boosted dark matter in noble liquid direct detection,''
Phys. Rev. D \textbf{99}, no.10, 103003 (2019)
%doi:10.1103/PhysRevD.99.103003
[arXiv:1812.05102 [hep-ph]].
%22 citations counted in INSPIRE as of 09 Sep 2021



%\cite{Kong:2014mia}
\bibitem{Kong:2014mia}
K.~Kong, G.~Mohlabeng and J.~C.~Park,
%``Boosted dark matter signals uplifted with self-interaction,''
Phys. Lett. B \textbf{743}, 256-266 (2015)
%doi:10.1016/j.physletb.2015.02.057
[arXiv:1411.6632 [hep-ph]].
%55 citations counted in INSPIRE as of 02 Jan 2022

%\cite{Kopp:2015bfa}
\bibitem{Kopp:2015bfa}
J.~Kopp, J.~Liu and X.~P.~Wang,
%``Boosted Dark Matter in IceCube and at the Galactic Center,''
JHEP \textbf{04}, 105 (2015)
%doi:10.1007/JHEP04(2015)105
[arXiv:1503.02669 [hep-ph]].
%82 citations counted in INSPIRE as of 02 Jan 2022

%\cite{Alhazmi:2016qcs}
\bibitem{Alhazmi:2016qcs}
H.~Alhazmi, K.~Kong, G.~Mohlabeng and J.~C.~Park,
%``Boosted Dark Matter at the Deep Underground Neutrino Experiment,''
JHEP \textbf{04}, 158 (2017)
%doi:10.1007/JHEP04(2017)158
[arXiv:1611.09866 [hep-ph]].
%34 citations counted in INSPIRE as of 02 Jan 2022

%\cite{Kim:2016zjx}
\bibitem{Kim:2016zjx}
D.~Kim, J.~C.~Park and S.~Shin,
%``Dark Matter \textquotedblleft{}Collider\textquotedblright{} from Inelastic Boosted Dark Matter,''
Phys. Rev. Lett. \textbf{119}, no.16, 161801 (2017)
%doi:10.1103/PhysRevLett.119.161801
[arXiv:1612.06867 [hep-ph]].
%43 citations counted in INSPIRE as of 30 Aug 2021

%\cite{Aoki:2018gjf}
\bibitem{Aoki:2018gjf}
M.~Aoki and T.~Toma,
%``Boosted Self-interacting Dark Matter in a Multi-component Dark Matter Model,''
JCAP \textbf{10}, 020 (2018)
%doi:10.1088/1475-7516/2018/10/020
[arXiv:1806.09154 [hep-ph]].
%20 citations counted in INSPIRE as of 30 Aug 2021

%\cite{Kim:2019had}
\bibitem{Kim:2019had}
D.~Kim, J.~C.~Park and S.~Shin,
%``Searching for boosted dark matter via dark-photon bremsstrahlung,''
Phys. Rev. D \textbf{100}, no.3, 035033 (2019)
%doi:10.1103/PhysRevD.100.035033
[arXiv:1903.05087 [hep-ph]].
%7 citations counted in INSPIRE as of 09 Sep 2021


%\cite{Kim:2018veo}
\bibitem{Kim:2018veo}
D.~Kim, K.~Kong, J.~C.~Park and S.~Shin,
%``Boosted Dark Matter Quarrying at Surface Neutrino Detectors,''
JHEP \textbf{08}, 155 (2018)
%doi:10.1007/JHEP08(2018)155
[arXiv:1804.07302 [hep-ph]].
%23 citations counted in INSPIRE as of 02 Jan 2022

%\cite{McKeen:2020vpf}
\bibitem{McKeen:2020vpf}
D.~McKeen, M.~Pospelov and N.~Raj,
%``Hydrogen Portal to Exotic Radioactivity,''
Phys. Rev. Lett. \textbf{125}, no.23, 231803 (2020)
%doi:10.1103/PhysRevLett.125.231803
[arXiv:2006.15140 [hep-ph]].
%41 citations counted in INSPIRE as of 02 Jan 2022


%\cite{Belanger:2013oya}
\bibitem{Belanger:2013oya}
G.~Belanger, F.~Boudjema, A.~Pukhov and A.~Semenov,
%``micrOMEGAs$\_$3: A program for calculating dark matter observables,''
Comput. Phys. Commun. \textbf{185}, 960-985 (2014)
%doi:10.1016/j.cpc.2013.10.016
[arXiv:1305.0237 [hep-ph]].
%673 citations counted in INSPIRE as of 04 Sep 2021

%\cite{Baratella:2013fya}
\bibitem{Baratella:2013fya}
P.~Baratella, M.~Cirelli, A.~Hektor, J.~Pata, M.~Piibeleht and A.~Strumia,
%``PPPC 4 DM$\nu$: a Poor Particle Physicist Cookbook for Neutrinos from Dark Matter annihilations in the Sun,''
JCAP \textbf{03}, 053 (2014)
%doi:10.1088/1475-7516/2014/03/053
[arXiv:1312.6408 [hep-ph]].
%67 citations counted in INSPIRE as of 04 Sep 2021


%\cite{Busoni:2013kaa}
\bibitem{Busoni:2013kaa}
G.~Busoni, A.~De Simone and W.~C.~Huang,
%``On the Minimum Dark Matter Mass Testable by Neutrinos from the Sun,''
JCAP \textbf{07}, 010 (2013)
%doi:10.1088/1475-7516/2013/07/010
[arXiv:1305.1817 [hep-ph]].
%61 citations counted in INSPIRE as of 06 Sep 2021


%\cite{Garani:2017jcj}
\bibitem{Garani:2017jcj}
R.~Garani and S.~Palomares-Ruiz,
%``Dark matter in the Sun: scattering off electrons vs nucleons,''
JCAP \textbf{05}, 007 (2017)
%doi:10.1088/1475-7516/2017/05/007
[arXiv:1702.02768 [hep-ph]].
%51 citations counted in INSPIRE as of 03 Jan 2022


%\cite{XENON:2019rxp}
\bibitem{XENON:2019rxp}
E.~Aprile \textit{et al.} [XENON],
%``Constraining the spin-dependent WIMP-nucleon cross sections with XENON1T,''
Phys. Rev. Lett. \textbf{122}, no.14, 141301 (2019)
%doi:10.1103/PhysRevLett.122.141301
[arXiv:1902.03234 [astro-ph.CO]].
%151 citations counted in INSPIRE as of 30 Aug 2021

%\cite{PandaX-II:2018woa}
\bibitem{PandaX-II:2018woa}
J.~Xia \textit{et al.} [PandaX-II],
%``PandaX-II Constraints on Spin-Dependent WIMP-Nucleon Effective Interactions,''
Phys. Lett. B \textbf{792}, 193-198 (2019)
%doi:10.1016/j.physletb.2019.02.043
[arXiv:1807.01936 [hep-ex]].
%47 citations counted in INSPIRE as of 09 Sep 2021


%\cite{PICO:2017tgi}
\bibitem{PICO:2017tgi}
C.~Amole \textit{et al.} [PICO],
%``Dark Matter Search Results from the PICO-60 C$_3$F$_8$ Bubble Chamber,''
Phys. Rev. Lett. \textbf{118}, no.25, 251301 (2017)
%doi:10.1103/PhysRevLett.118.251301
[arXiv:1702.07666 [astro-ph.CO]].
%320 citations counted in INSPIRE as of 03 Sep 2021

%\cite{IceCube:2016yoy}
\bibitem{IceCube:2016yoy}
M.~G.~Aartsen \textit{et al.} [IceCube],
%``Improved limits on dark matter annihilation in the Sun with the 79-string IceCube detector and implications for supersymmetry,''
JCAP \textbf{04}, 022 (2016)
%doi:10.1088/1475-7516/2016/04/022
[arXiv:1601.00653 [hep-ph]].
%185 citations counted in INSPIRE as of 03 Sep 2021




%\cite{Bertone:2009cb}
\bibitem{Bertone:2009cb}
G.~Bertone, C.~B.~Jackson, G.~Shaughnessy, T.~M.~P.~Tait and A.~Vallinotto,
%``The WIMP Forest: Indirect Detection of a Chiral Square,''
Phys. Rev. D \textbf{80}, 023512 (2009)
%doi:10.1103/PhysRevD.80.023512
[arXiv:0904.1442 [astro-ph.HE]].
%79 citations counted in INSPIRE as of 04 Sep 2021

%\cite{Ibarra:2012dw}
\bibitem{Ibarra:2012dw}
A.~Ibarra, S.~Lopez Gehler and M.~Pato,
%``Dark matter constraints from box-shaped gamma-ray features,''
JCAP \textbf{07}, 043 (2012)
%doi:10.1088/1475-7516/2012/07/043
[arXiv:1205.0007 [hep-ph]].
%136 citations counted in INSPIRE as of 04 Jan 2022


%\cite{Super-Kamiokande:2015qek}
\bibitem{Super-Kamiokande:2015qek}
E.~Richard \textit{et al.} [Super-Kamiokande],
%``Measurements of the atmospheric neutrino flux by Super-Kamiokande: energy spectra, geomagnetic effects, and solar modulation,''
Phys. Rev. D \textbf{94}, no.5, 052001 (2016)
%doi:10.1103/PhysRevD.94.052001
[arXiv:1510.08127 [hep-ex]].
%89 citations counted in INSPIRE as of 05 Sep 2021


%\cite{Honda:2011nf}
\bibitem{Honda:2011nf}
M.~Honda, T.~Kajita, K.~Kasahara and S.~Midorikawa,
%``Improvement of low energy atmospheric neutrino flux calculation using the JAM nuclear interaction model,''
Phys. Rev. D \textbf{83}, 123001 (2011)
%doi:10.1103/PhysRevD.83.123001
[arXiv:1102.2688 [astro-ph.HE]].
%202 citations counted in INSPIRE as of 09 Sep 2021

%\cite{Berger:2019ttc}
\bibitem{Berger:2019ttc}
J.~Berger, Y.~Cui, M.~Graham, L.~Necib, G.~Petrillo, D.~Stocks, Y.~T.~Tsai and Y.~Zhao,
%``Prospects for detecting boosted dark matter in DUNE through hadronic interactions,''
Phys. Rev. D \textbf{103}, no.9, 095012 (2021)
%doi:10.1103/PhysRevD.103.095012
[arXiv:1912.05558 [hep-ph]].
%7 citations counted in INSPIRE as of 06 Sep 2021


%\cite{DUNE:2015lol}
\bibitem{DUNE:2015lol}
R.~Acciarri \textit{et al.} [DUNE],
%``Long-Baseline Neutrino Facility (LBNF) and Deep Underground Neutrino Experiment (DUNE): Conceptual Design Report, Volume 2: The Physics Program for DUNE at LBNF,''
[arXiv:1512.06148 [physics.ins-det]].
%763 citations counted in INSPIRE as of 07 Sep 2021


%\cite{MAGIC:2016xys}
\bibitem{MAGIC:2016xys}
M.~L.~Ahnen \textit{et al.} [MAGIC and Fermi-LAT],
%``Limits to Dark Matter Annihilation Cross-Section from a Combined Analysis of MAGIC and Fermi-LAT Observations of Dwarf Satellite Galaxies,''
JCAP \textbf{02}, 039 (2016)
%doi:10.1088/1475-7516/2016/02/039
[arXiv:1601.06590 [astro-ph.HE]].
%307 citations counted in INSPIRE as of 04 Sep 2021


%\cite{Gelmini:2018ogy}
\bibitem{Gelmini:2018ogy}
G.~B.~Gelmini, V.~Takhistov and S.~J.~Witte,
%``Casting a Wide Signal Net with Future Direct Dark Matter Detection Experiments,''
JCAP \textbf{07}, 009 (2018)
[erratum: JCAP \textbf{02}, E02 (2019)]
%doi:10.1088/1475-7516/2018/07/009
[arXiv:1804.01638 [hep-ph]].
%23 citations counted in INSPIRE as of 02 Sep 2021



%\cite{Ma:2007gq}
\bibitem{Ma:2007gq}
E.~Ma,
%``Z(3) Dark Matter and Two-Loop Neutrino Mass,''
Phys. Lett. B \textbf{662}, 49-52 (2008)
%doi:10.1016/j.physletb.2008.02.053
[arXiv:0708.3371 [hep-ph]].
%95 citations counted in INSPIRE as of 30 Aug 2021

%\cite{Aoki:2014cja}
\bibitem{Aoki:2014cja}
M.~Aoki and T.~Toma,
%``Impact of semi-annihilation of $\mathbb{Z}_3$ symmetric dark matter with radiative neutrino masses,''
JCAP \textbf{09}, 016 (2014)
%doi:10.1088/1475-7516/2014/09/016
[arXiv:1405.5870 [hep-ph]].
%61 citations counted in INSPIRE as of 30 Aug 2021


%\cite{Ho:2016aye}
\bibitem{Ho:2016aye}
S.~Y.~Ho, T.~Toma and K.~Tsumura,
%``Systematic $U(1)_{B-L}$ extensions of loop-induced neutrino mass models with dark matter,''
Phys. Rev. D \textbf{94}, no.3, 033007 (2016)
%doi:10.1103/PhysRevD.94.033007
[arXiv:1604.07894 [hep-ph]].
%22 citations counted in INSPIRE as of 30 Aug 2021

\end{thebibliography}
\end{document}